\documentclass[reprint, 
 amsmath,amssymb,
 aps, physrev,
]{revtex4-2}

\usepackage{float}
\usepackage{graphicx}% Include figure files
\usepackage{dcolumn}% Align table columns on decimal point
\usepackage{bm}% bold math
\usepackage{hyperref}% add hypertext capabilities
\usepackage{xcolor}
\begin{document}

%\preprint{APS/123-QED}

\title{\textbf{Scalar propagator in a rotating thermal medium: Consistency checks and applications} 
}% 

\author{Luis A. Hernández}
 \affiliation{Departamento de F\'isica, Universidad Aut\'onoma Metropolitana-Iztapalapa, Avenida San Rafael Atlixco 186, Ciudad de México 09340, Mexico.}%Lines break automatically or can be forced with \\
\author{R. Zamora}
\email[Corresponding author: ]{rrzamora@uc.cl}
\affiliation{Facultad de Ingenier\'ia y Arquitectura, Universidad Central de Chile, Av. Sta. Isabel 1186, Santiago 8330601, Chile}
\affiliation{Facultad de Ingenier\'ia, Universidad San Sebasti\'an, Bellavista 7, Recoleta, Santiago, Chile.}

\date{\today}% It is always \today, today,

\begin{abstract}
We derive the scalar propagator in a thermal medium under global rotation without restricting the spacetime points to a common comoving trajectory. The resulting propagator retains an explicit dependence on the radial coordinates, reflecting the lack of translational invariance in the transverse plane, and therefore cannot in general be formulated solely in momentum space. We show that the usual free Feynman propagator is exactly recovered in the nonrotating limit, providing a first consistency check of the formulation. As a nontrivial test, we apply the propagator to the interacting $\lambda\phi^4$ theory at finite temperature and rotation. The one-loop scalar self-energy obtained within the Matsubara formalism correctly separates into vacuum and medium contributions and reproduces the standard finite-temperature result when $\Omega\to0$. We subsequently construct the effective potential including screening effects through ring resummation and, for $\Omega/T\ll1$, evaluate the rotational corrections up to $\mathcal{O}(\Omega^4)$. Within this model, the resulting effective potential exhibits the expected thermal restoration of the spontaneously broken $\mathbb{Z}_2$ symmetry, while global rotation favors symmetry restoration and enhances this effect as the transverse size of the system increases. These results provide complementary consistency checks of the rotating scalar propagator and highlight the need to retain its explicit position-space structure in more general perturbative calculations.
\end{abstract}

\maketitle

%\tableofcontents

\section{Introduction \label{Sec1}}

Relativistic heavy-ion collisions provide a unique experimental environment to investigate strongly interacting matter under extreme conditions~\cite{Jacobs:2004qv,Sorin:2011zz,CBM:2016kpk,ALICE:2022wpn,Chen:2024aom,Shou:2024uga}. Over the past decades, an increasingly detailed understanding of the dynamics of these collisions has revealed that the produced medium is characterized not only by extremely high temperatures, but also by exceptionally intense electromagnetic fields~\cite{Skokov:2009qp,Voronyuk:2011jd,Tuchin:2013ie,Brandenburg:2021lnj,STAR:2023jdd,Taya:2024wrm,Panda:2024ccj,Adhikari:2024bfa,Mustafa:2025uad}, large baryon densities~\cite{Klahn:2006ir,Randrup:2006nr,Randrup:2009ch,Fukushima:2013rx,Steinheimer:2025hsr}, and enormous angular velocities~\cite{Becattini:2015ska,Fukushima:2018grm,Xia:2018tes,Ivanov:2019wzg,Becattini:2020ngo,Tsegelnik:2022eoz}, reaching values among the largest known in the Universe. The simultaneous presence of these extreme conditions introduces additional scales and nontrivial effects into the evolution of the system, substantially enriching the dynamics of the collision. From a theoretical perspective, understanding how strongly interacting matter responds to such extreme environments therefore represents a formidable challenge and has become a subject of sustained interest in high-energy nuclear physics.

In this work, we focus on the interplay between thermal and vortical effects, leaving aside the additional influence of electromagnetic fields and finite baryon density. The remarkably large angular momentum carried by noncentral relativistic heavy-ion collisions~\cite{Csernai:2014ywa,Deng:2016gyh,STAR:2017ckg,Deng:2020ygd} can be partially transferred to the produced medium, giving rise to vortical motion and providing a unique setting in which the response of quantum fields to rotation can be investigated. From the theoretical point of view, however, incorporating rotation consistently into finite-temperature quantum field theory requires particular care, since the rotational background modifies the field modes and, consequently, the structure of the corresponding propagators. Different representations of the scalar and fermion propagators in a globally rotating background have been employed in the literature~\cite{Vilenkin:1980zv,Gaspar:2023nqk,Siri:2024scq,Gecic:2026iry,WeiMingHua:2020eee,Ayala:2021osy,Castano-Yepes:2025zae,Kawaguchi:2025mkh}. In this work, we revisit this problem and derive the scalar propagator without imposing the comoving-frame restriction adopted in our previous formulation~\cite{Gaspar:2023nqk,Hernandez:2024nev}. We show that the resulting expression provides the consistent propagator for a scalar field in a rotating thermal medium, as demonstrated by its consistency with the nonrotating limit and, importantly, by its ability to recover the vacuum contribution once the Matsubara formalism is implemented. As a nontrivial application of this formulation, we employ the interacting $\lambda\phi^4$ theory and construct its effective potential at finite temperature and global rotation, including ring-diagram contributions to account for collective screening effects. The resulting framework is then used to investigate the influence of global rotation on the symmetry-restoration phase transition.

The remainder of this work is organized as follows. In Sec.~\ref{Sec2}, we derive the scalar-field propagator in a globally rotating background without restricting the description to a comoving reference frame, and discuss the main consistency properties of the resulting expression. In Sec.~\ref{Sec3}, we apply this propagator to the interacting $\lambda\phi^4$ theory to investigate symmetry restoration in a system subject simultaneously to finite temperature and global rotation. To this end, we construct the effective potential including ring-diagram contributions, which requires the evaluation of the scalar self-energy at finite $T$ and $\Omega$ and allows us to consistently incorporate screening effects. In Sec.~\ref{Sec4}, we present our numerical results and discuss the impact of rotation and screening on the symmetry-restoration phase transition. Finally, in Sec.~\ref{Sec5}, we summarize our main findings and present our conclusions.

%%%%%%%%%%%%%%%%%%%%%%%%%%%%%%%%%%%%%%%%%%%%%%%%%%%%%%%
\section{Scalar propagator in a rotating background \label{Sec2}}

We consider a relativistic scalar field in a system undergoing rigid rotation around the $z$ axis with constant angular velocity $\Omega$. As in our previous work~\cite{Gaspar:2023nqk}, the rotating system is described as a cylinder of radius $R$, with the causality condition
\begin{equation}
\Omega R<1.
\end{equation}
The scalar propagator can be constructed from the eigenfunctions of the Klein–Gordon operator in the rotating background. In contrast to our previous treatment, however, we do not restrict the two spacetime points entering the propagator to belong to the same comoving trajectory. As we show below, retaining their independent spacetime coordinates is essential for obtaining the appropriate propagator and its thermal extension.

For a scalar field of mass $m$, the Klein–Gordon equation in the rotating frame can be written as
\begin{align}
\bigg[\Big(&i\frac{\partial}{\partial t}+\Omega \hat L_z\Big)^2+\frac{\partial^2}{\partial r^2}+\frac{1}{r}\frac{\partial}{\partial r}\nonumber \\
&+\frac{1}{r^2}\frac{\partial^2}{\partial\varphi^2}+\frac{\partial^2}{\partial z^2}-m^2\Big]\Phi(x)=0,
\label{KGrotation}
\end{align}
where
\begin{equation}
\hat L_z=-i\frac{\partial}{\partial\varphi}
\end{equation}
is the angular-momentum operator along the rotation axis. Owing to the cylindrical symmetry of the problem, the solutions can be written as
\begin{equation}
\Phi_{\ell}(x)=e^{-ip_0t+ip_z z+i\ell\varphi}J_{\ell}(p_\perp r),
\label{ScalarModes}
\end{equation}
where $\ell\in\mathbb{Z}$ is the azimuthal angular-momentum quantum number and $J_\ell$ denotes the Bessel function of the first kind. Substitution of Eq.\eqref{ScalarModes} into Eq.\eqref{KGrotation} leads to the corresponding eigenvalue
\begin{equation}
(p_0+\ell\Omega)^2-p_z^2-p_\perp^2-m^2.
\label{ScalarEigenvalue}
\end{equation}

Using the spectral representation of the Green’s function in terms of these modes, the scalar propagator can therefore be written as
\begin{align}
D(x,x')&=\sum_{\ell=-\infty}^{\infty}\int\frac{dp_0 dp_z p_\perp dp_\perp}{(2\pi)^3}\nonumber \\
&\times \frac{e^{-ip_0(t-t')}e^{ip_z(z-z')}e^{i\ell(\varphi-\varphi')}}{(p_0+\ell\Omega)^2-p_z^2-p_\perp^2-m^2+i\epsilon} \nonumber\\
&\times J_\ell(p_\perp r)J_\ell(p_\perp r').
\label{ScalarPropagatorRotation}
\end{align}
Equation~(\ref{ScalarPropagatorRotation}) constitutes the starting point of our analysis. Notice that the rotational effect is encoded through the angular-momentum-dependent energy shift $p_0\rightarrow p_0+\ell\Omega$, while the complete dependence on the two spacetime points is retained. In particular, no relation between $\varphi-\varphi'$ and $t-t'$ has been imposed. As a consequence, unlike in our previous treatment, the coordinate dependence of the propagator cannot be completely eliminated in favor of a momentum-space representation.

An important consequence of keeping the two spacetime points independent should be emphasized. In our previous treatment, the condition
\begin{equation}
\Omega=\frac{\varphi-\varphi'}{t-t'}
\end{equation}
was imposed by assuming that both points followed the same comoving trajectory. This relation considerably simplifies the coordinate dependence of the Green’s function and ultimately allows the propagator to be expressed solely in momentum space. Here, no such condition is imposed. Consequently, the propagator retains an explicit dependence on the spatial coordinates through the radial functions $J_\ell(p_\perp r)J_\ell(p_\perp r')$, reflecting the lack of translational invariance in the transverse plane induced by the rotating background.

This feature modifies the way loop contributions must be constructed. In a translationally invariant system, the coordinate dependence can be completely factorized from a closed loop, so that the spacetime integration produces an overall spacetime-volume factor and the corresponding density can subsequently be expressed entirely in momentum space. In the present case, this factorization is only partial. Loop contributions must therefore be evaluated by integrating both over the momentum modes and over the spacetime region occupied by the rotating system. For rigid rotation around the $z$ axis, we consider
\begin{align}
-\infty&<t<\infty,\qquad
-\infty<z<\infty, \nonumber \\
0&\leq\varphi<2\pi,\qquad
0\leq r\leq R,
\end{align}
where the cylinder radius is restricted by the causality condition
\begin{equation}
\Omega R<1.
\end{equation}
Accordingly,
\begin{equation}
d^4x=dt\, dz\, d\varphi\, rdr.
\end{equation}

For example, in the coincident limit
\begin{align}
D(x,x)&=
\sum_{\ell=-\infty}^{\infty}
\int
\frac{dp_0 dp_z p_\perp dp_\perp}{(2\pi)^3} \nonumber \\
&\times \frac{J_\ell^2(p_\perp r)}
{(p_0+\ell\Omega)^2-p_z^2-p_\perp^2-m^2+i\epsilon},
\label{CoincidentPropagator}
\end{align}
so that the integrations over $t$, $z$, and $\varphi$ factorize, whereas the radial integration remains coupled to the momentum dependence through the Bessel functions. The corresponding loop density is obtained only after carrying out this remaining spatial integration and dividing the full loop contribution by the spacetime volume of the cylindrical system. Thus, unlike the translationally invariant case, the calculation cannot be formulated from the outset solely in momentum space.

As a first consistency check of Eq.~\eqref{ScalarPropagatorRotation},
we consider the nonrotating limit, $\Omega\to 0$. In this limit, the
denominator becomes independent of the angular-momentum quantum number
$\ell$, and the propagator reduces to
\begin{align}
D(x,x')\Big|_{\Omega=0}
&=
\int
\frac{dp_0\,dp_z\,p_\perp dp_\perp}{(2\pi)^3} \nonumber \\
&\times \frac{e^{-ip_0(t-t')}e^{ip_z(z-z')}}{p_0^2-p_z^2-p_\perp^2-m^2+i\epsilon} \nonumber\\
&\times \sum_{\ell=-\infty}^{\infty}e^{i\ell(\varphi-\varphi')}J_\ell(p_\perp r)J_\ell(p_\perp r').
\label{PropagatorOmegaZero}
\end{align}
The sum over the angular-momentum modes can now be performed using the Bessel addition theorem,
\begin{equation}
\sum_{\ell=-\infty}^{\infty}e^{i\ell(\varphi-\varphi')}J_\ell(p_\perp r)J_\ell(p_\perp r')=J_0\left(p_\perp
\left|\boldsymbol{x}_\perp-\boldsymbol{x}'_\perp\right|\right),
\label{BesselAddition}
\end{equation}
where
\begin{equation}
\left|\boldsymbol{x}_\perp-\boldsymbol{x}'_\perp\right|=\sqrt{r^2+r'^2-2rr'\cos(\varphi-\varphi')}.
\end{equation}
Therefore, Eq.~\eqref{PropagatorOmegaZero} becomes
\begin{align}
D(x,x')\Big|_{\Omega=0}
&=\int\frac{dp_0\,dp_z\,p_\perp dp_\perp}{(2\pi)^3}J_0\left(p_\perp\left|\boldsymbol{x}_\perp-\boldsymbol{x}'_\perp\right|\right) \nonumber \\
&\times\frac{e^{-ip_0(t-t')}e^{ip_z(z-z')}}{p_0^2-p_z^2-p_\perp^2-m^2+i\epsilon}.
\label{FreeScalarPropagatorCylindrical}
\end{align}
The separate dependence on the two transverse coordinates has thus
combined into a dependence only on their relative separation,
$\boldsymbol{x}_\perp-\boldsymbol{x}'_\perp$, showing explicitly that
translational invariance is recovered in the nonrotating limit.
Equation~\eqref{FreeScalarPropagatorCylindrical} is precisely the free Feynman propagator expressed in cylindrical coordinates. To make this connection explicit, we use the integral representation
\begin{equation}
J_0\left(p_\perp
\left|\boldsymbol{x}_\perp-\boldsymbol{x}'_\perp\right|\right)
=
\frac{1}{2\pi}
\int_0^{2\pi}d\phi_p\,
e^{i\boldsymbol{p}_\perp\cdot
(\boldsymbol{x}_\perp-\boldsymbol{x}'_\perp)}.
\label{J0IntegralRepresentation}
\end{equation}
Since
\begin{equation}
d^2p_\perp=p_\perp dp_\perp d\phi_p,
\end{equation}
Eq.~\eqref{FreeScalarPropagatorCylindrical} can be rewritten as
\begin{equation}
D(x,x')\Big|_{\Omega=0}
=
\int\frac{d^4p}{(2\pi)^4}
\frac{e^{-ip\cdot(x-x')}}
{p^2-m^2+i\epsilon},
\label{FreeFeynmanPropagator}
\end{equation}
which is the usual free Feynman propagator. Therefore, the
nonrotating limit not only removes the angular-momentum-dependent energy shift, but also restores the translational invariance of the vacuum propagator.

We have thus obtained a scalar propagator that consistently
incorporates the effects of global rotation while retaining the spatial dependence associated with the rotating background. Unlike in a translationally invariant system, this dependence prevents a complete formulation in momentum space and must be explicitly accounted for when constructing loop contributions. At the same time, the exact recovery of the usual Feynman propagator in the limit $\Omega\to0$ provides a first consistency check of the present formulation.

A further and more stringent test is provided by using this propagator in actual loop calculations. Rather than introducing a more elaborate model, our purpose in the following section is to employ the simple interacting $\lambda\phi^4$ theory as a controlled framework in which the consistency of the rotating propagator can be explicitly examined. In particular, we calculate the scalar self-energy and the effective potential at finite temperature and rotation, and verify that the Matsubara formalism yields the expected separation between vacuum and medium contributions. The same calculation also allows us to
incorporate screening effects and study symmetry restoration as a physical application of the resulting framework.

%%%%%%%%%%%%%%%%%%%%%%%%%%%%%%%%%%%%%%%%%%%%%%%%%%%%%%%
\section{Effective Potential and Screening in the $\lambda\phi^4$ Theory \label{Sec3}}

We now apply the scalar propagator derived in the previous section to an interacting scalar theory with spontaneous symmetry breaking. We consider the $\lambda\phi^4$ model described by the Lagrangian
\begin{equation}
\mathcal{L}=\frac{1}{2}(\partial_\mu\phi)^2+\frac{a^2}{2}\phi^2-\frac{\lambda}{4}\phi^4,
\label{LagrangianPhi4}
\end{equation}
where $\phi$ is a real scalar field, $a^2>0$ is the squared mass parameter, and $\lambda>0$ is the scalar self-coupling. The theory is invariant under the discrete transformation
\begin{equation}
\phi\rightarrow-\phi,
\end{equation}
corresponding to a $\mathbb{Z}_2$ symmetry. The tree-level potential is
\begin{equation}
V_{\rm tree}(\phi)=-\frac{a^2}{2}\phi^2+\frac{\lambda}{4}\phi^4,
\label{TreeLevelPotential}
\end{equation}
which develops two degenerate minima and therefore allows for spontaneous symmetry breaking.

To describe the broken-symmetry phase, we shift the scalar field according to
\begin{equation}
\phi\rightarrow\phi+v,
\label{ScalarShift}
\end{equation}
where $v$ denotes the vacuum expectation value of the scalar field and plays the role of the order parameter. After the shift, the Lagrangian becomes
\begin{align}
\mathcal{L}&=\frac{1}{2}(\partial_\mu\phi)^2-\frac{1}{2}m^2\phi^2-\lambda v\phi^3-\frac{\lambda}{4}\phi^4 \nonumber\\
&+\frac{a^2}{2}v^2-\frac{\lambda}{4}v^4+(a^2v-\lambda v^3)\phi,
\label{ShiftedLagrangian}
\end{align}
where the field-dependent scalar mass is
\begin{equation}
m^2=3\lambda v^2-a^2.
\label{ScalarMass}
\end{equation}
At tree level, the stationary condition,
\begin{equation}
\frac{dV_{\rm tree}(v)}{dv}=0,
\end{equation}
gives the nonvanishing vacuum expectation values
\begin{equation}
v_0=\pm\sqrt{\frac{a^2}{\lambda}},
\label{TreeVEV}
\end{equation}
which characterize the spontaneously broken phase. The restoration of the $\mathbb{Z}_2$ symmetry at finite temperature and global rotation can therefore be studied by following the evolution of the minimum of the effective potential as a function of $T$ and $\Omega$.

We first construct the effective potential up to one-loop order,
\begin{equation}
V_{\rm eff}^{(1)}(v;T,\Omega)=V_{\rm tree}(v)+V_{1}(v;T,\Omega),
\label{VeffOneLoop}
\end{equation}
where $V_{\rm tree}$ is given in Eq.~\eqref{TreeLevelPotential}, while $V_{1}$ contains the quantum and medium corrections generated by the scalar fluctuations. At this order, the thermal contribution is built from noninteracting quasiparticle modes propagating in the rotating background. Therefore, although the one-loop term incorporates the effects of finite temperature and global rotation through the modified scalar propagator, it does not yet account for collective interactions among the excitations of the medium. In this sense, the one-loop approximation retains an essentially ideal-gas-like structure for the thermal degrees of freedom.

To go beyond this approximation and incorporate collective screening effects, we include the resummation of the ring, or daisy, diagrams. These contributions account for the modification of the scalar propagation induced by the medium through the corresponding self-energy and are particularly relevant in the vicinity of a phase transition, where infrared effects become increasingly important. The effective potential is then written as
\begin{equation}
V_{\rm eff}(v;T,\Omega)=V_{\rm tree}(v)+V_{1}(v;T,\Omega)+V_{\rm ring}(v;T,\Omega).
\label{VeffRings}
\end{equation}
The ring contribution is constructed by resumming insertions of the scalar self-energy $\Pi(T,\Omega)$ in the bosonic propagator. Besides incorporating plasma-screening effects and thereby going beyond the one-loop approximation, this resummation also improves the infrared behavior of the effective potential. In particular, when the field-dependent squared mass, Eq.~(\ref{ScalarMass}), becomes negative in part of the field space, the one-loop potential develops nonanalytic contributions associated with the appearance of tachyonic modes. The ring resummation reorganizes these infrared-sensitive terms by replacing the bare field-dependent mass with a screened mass,
\begin{equation}
m^2\longrightarrow m^2+\Pi(T,\Omega),
\end{equation}
thereby providing a consistent treatment of the effective potential in the region relevant for symmetry restoration.

Since the inclusion of ring diagrams requires the scalar self-energy, it is convenient to determine this quantity first. We therefore compute $\Pi(T,\Omega)$ using the rotating scalar propagator derived in Sec.~\ref{Sec2}. The resulting self-energy will subsequently enter the ring resummation through the screened scalar mass $m^2+\Pi(T,\Omega)$.

At one-loop order, the scalar self-energy receives a contribution from the tadpole diagram. For the interaction term $-\lambda\phi^4/4$, its initial expression is
\begin{equation}
    -i\Pi=-\frac{i\lambda}{4}(12)\int d^4x e^{ip\cdot (x-x)}iD(x,x),
    \label{Selfenergyinitial}
\end{equation}
where the factor $12$ accounts for the possible contractions of the scalar fields. Substituting the coincident propagator, Eq.~\eqref{CoincidentPropagator}, into Eq.~\eqref{Selfenergyinitial}, we obtain
\begin{align}
    -i\Pi=-\frac{i\lambda}{4}(12)&\int d^4x \, i\sum_{\ell=-\infty}^{\infty}\int\frac{dp_0 dp_z p_\perp dp_\perp}{(2\pi)^3} \nonumber \\
&\times \frac{J_\ell^2(p_\perp r)}
{(p_0+\ell\Omega)^2-p_z^2-p_\perp^2-m^2+i\epsilon}.
\label{Selfenergyinitialexplicit}
\end{align}

As discussed in Sec.~\ref{Sec2}, the rotating propagator retains an explicit dependence on the radial coordinate. Consequently, the spacetime integration cannot be completely factorized before evaluating the loop. To obtain the self-energy density, we divide Eq.~\eqref{Selfenergyinitialexplicit} by the spacetime volume of the cylindrical system,
\begin{equation}
    \mathcal{V}=\int dt\,dz\,d\varphi \int_0^R r\,dr=\frac{R^2}{2}\int dt\,dz\,d\varphi,
    \label{SpacetimeVolume}
\end{equation}
where the finite radial domain is restricted by the causality
condition $\Omega R<1$.

We next perform a Wick rotation and implement the imaginary-time formalism to incorporate finite-temperature effects. The energy integral is replaced by a sum over bosonic Matsubara frequencies, $\omega_n=2\pi nT$, yielding
\begin{align}
    \Pi(T,\Omega)&=\frac{3\lambda}{\mathcal{V}}\int d^4x \, T\sum_{n=-\infty}^\infty\sum_{\ell=-\infty}^{\infty}\int\frac{dp_z \, p_\perp dp_\perp}{(2\pi)^2}\nonumber \\
&\times \frac{J_\ell^2(p_\perp r)}
{(\omega_n-i\ell\Omega)^2+p_z^2+p_\perp^2+m^2}\nonumber \\
&=\frac{6\lambda}{R^2}\int_0^R rdr\,T\sum_{n=-\infty}^\infty\sum_{\ell=-\infty}^{\infty}\int\frac{dp_z \, p_\perp dp_\perp}{(2\pi)^2}\nonumber \\
&\times \frac{J_\ell^2(p_\perp r)}
{(\omega_n-i\ell\Omega)^2+p_z^2+p_\perp^2+m^2},
\end{align} 
Performing the sum over the Matsubara modes, we find
\begin{align}
   \Pi(T,\Omega)&= \frac{6\lambda}{R^2}\int_0^R rdr\, \sum_{\ell=-\infty}^{\infty}\int\frac{dp_z \, p_\perp dp_\perp}{(2\pi)^2}\nonumber \\
   &\times \frac{J_\ell^2(p_\perp r)}{2E_p}\big(1+n_B(E_p-\ell \Omega)+n_B(E_p+\ell \Omega) \big),
   \label{selfenergyafterMatsubarasum}
\end{align}
where $E_p\equiv\sqrt{p_z^2+p_\perp^2+m^2}$ and $n_B(x)$ denotes the Bose--Einstein distribution. Equation~(\ref{selfenergyafterMatsubarasum}) exhibits an important feature of the rotating propagator derived in Sec.~\ref{Sec2}. The term independent of the Bose--Einstein distributions corresponds to the vacuum contribution, whereas the remaining terms describe the medium contribution and contain the combined effects of temperature and rotation. Therefore, the Matsubara sum naturally recovers the expected separation of the self-energy into vacuum and matter contributions.

To proceed analytically, we consider the regime $\Omega/T\ll1$ and expand the Bose--Einstein distributions around $\Omega/T=0$. Since the combination entering Eq.~\eqref{selfenergyafterMatsubarasum} is symmetric under $\Omega\rightarrow-\Omega$, only even powers of $\Omega$ survive. Up to $\mathcal{O}[(\Omega/T)^4]$, we obtain
\begin{align}
    \Pi(T,\Omega)&= \frac{6\lambda}{R^2}\int_0^R rdr\, \sum_{\ell=-\infty}^{\infty}\int\frac{dp_z \, p_\perp dp_\perp}{(2\pi)^2}\nonumber \\
   &\times \frac{J_\ell^2(p_\perp r)}{2E_p}\bigg[1+2n_B(E_p)+\frac{\ell^2\Omega^2}{T^2}B_2(E_p)\nonumber \\
   &+\frac{\ell^4\Omega^4}{12T^4}B_4(E_p)+\mathcal{O}\left(\left(\Omega/T\right)^6\right) \bigg],
   \label{selfenergylowOmega}
\end{align}
where, for compactness, we have defined
\begin{align}
    B_2(E_p)&\equiv n_B(E_p)\left[1+n_B(E_p)\right]\left[1+2n_B(E_p)\right], \nonumber \\
    B_4(E_p)&\equiv n_B(E_p)\left[1+n_B(E_p)\right]\left[1+14n_B(E_p)\right.\nonumber \\
    &\left.+36n_B(E_p)^2+24n_B(E_P)^3\right].
\end{align}

The sum over the angular-momentum modes can now be carried out analytically by using the identities
\begin{align}
    \sum_{\ell=-\infty}^\infty
J_\ell^2(x)&=1, \nonumber \\
\sum_{\ell=-\infty}^\infty
\ell^2J_\ell^2(x)&=\frac{x^2}{2}, \nonumber \\
\sum_{\ell=-\infty}^\infty
\ell^4J_\ell^2(x)&=\frac{x^2}{2}+\frac{3x^4}{8}.
\label{sumangularmodes}
\end{align}
Substituting these relations into Eq.~\eqref{selfenergylowOmega}, we obtain
\begin{align}
    \Pi(T,\Omega)&= \frac{6\lambda}{R^2}\int_0^R rdr\, \int\frac{dp_z \, p_\perp dp_\perp}{(2\pi)^2}\nonumber \\
   &\times \frac{1}{2E_p}\Bigg[1+2n_B(E_p)+\frac{\Omega^2(p_\perp r)^2}{2T^2}B_2(E_p)\nonumber \\
   &+\frac{\Omega^4}{T^4}\left(\frac{(p_\perp r)^2}{24}+\frac{(p_\perp r)^4}{32}\right)B_4(E_p) \Bigg].
\end{align}
The remaining radial integration can be performed straightforwardly. Using
\begin{align}
    \frac{1}{R^2}\int_0^R rdr&=\frac{1}{2}, \nonumber \\
    \frac{1}{R^2}\int_0^R r^3dr&=\frac{R^2}{4}, \nonumber \\
    \frac{1}{R^2}\int_0^R r^5dr&=\frac{R^4}{6},
    \label{intradial}
\end{align}
we finally obtain
\begin{align}
    \Pi(T,\Omega)&=3\lambda \int\frac{dp_z \, p_\perp dp_\perp}{(2\pi)^2}\frac{1}{2E_p} \nonumber \\
    &\times \Bigg[ 1+2n_B(E_p)+\frac{\Omega^2(p_\perp R)^2}{4T^2}B_2(E_p) \nonumber \\
    &+\frac{\Omega^4}{T^4}\left( \frac{(p_\perp R)^2}{48}+ \frac{(p_\perp R)^4}{96} \right)B_4(E_p)\Bigg].
    \label{selfenergyfinalexpression}
\end{align}
From Eq.~(\ref{selfenergyfinalexpression}), taking the limit $\Omega\to0$ reproduces the well-known expression for the self-energy of a neutral scalar field at finite temperature, including both its vacuum and medium contributions. Importantly, the same result can be obtained without invoking the $\Omega/T\ll1$ expansion. Indeed, one may set $\Omega=0$ directly in Eq.~(\ref{selfenergyafterMatsubarasum}) and use the first identity in Eq.~(\ref{sumangularmodes}), recovering the same expression. This provides an additional consistency check of the rotating propagator derived in this work.

Having determined the scalar self-energy, we can now use it to construct the one-loop contribution to the effective potential. Indeed, the tadpole self-energy is directly related to the derivative of $V_1$ with respect to the field-dependent squared mass. Taking into account the combinatorial factor associated with the $-\lambda\phi^4/4$ interaction, this relation can be written as
\begin{equation}
    \Pi(T,\Omega)=12\frac{\lambda}{4}\left(2\frac{dV_1}{dm^2}
    \right)=6\lambda\frac{dV_1}{dm^2}.
    \label{PiV1Relation}
\end{equation}
Therefore, the one-loop effective potential can be obtained directly from the self-energy derived above as
\begin{equation}
    V_1(v;T,\Omega)=\frac{1}{6\lambda}\int dm^2\,\Pi(T,\Omega),
    \label{V1FromSelfEnergy}
\end{equation}
up to an additive term independent of $m^2$. This procedure ensures that the one-loop effective potential is constructed consistently from the same rotating scalar propagator used in the self-energy calculation. 

Performing the integration over $m^2$ in Eq.~\eqref{selfenergyafterMatsubarasum}, we obtain
\begin{align}
    V_1(v;T,\Omega)&=\frac{2}{R^2}\int_0^R rdr \sum_{\ell=-\infty}^\infty\int \frac{dp_z \, p_\perp dp_\perp}{(2\pi)^2}J_\ell^2(p_\perp r) \nonumber \\
    &\times \frac{1}{2}\left[ E_p+T\ln (1-e^{-(E_p-\ell \Omega)/T}) \right. \nonumber \\
    &\left. + T\ln (1-e^{-(E_p+\ell \Omega)/T})\right].
    \label{V1initial}
\end{align}
This expression explicitly exhibits the expected separation between vacuum and medium contributions. The first term, proportional to $E_p$, corresponds to the zero-temperature vacuum contribution, whereas the logarithmic terms contain the effects of the thermal medium and global rotation. This separation parallels that found previously for the scalar self-energy and provides a further consistency check of the rotating propagator employed throughout this work.

As in the self-energy calculation, we consider the regime $\Omega/T\ll1$ and expand the rotationally dependent terms around $\Omega/T=0$. Since the two logarithmic contributions are related by $\Omega\rightarrow-\Omega$, their sum is an even function of $\Omega$, and therefore only even powers of the angular velocity survive. Up to fourth order in $\Omega$, we obtain
\begin{align}
    V_1(v;T,\Omega)&=\frac{2}{R^2}\int_0^R rdr \sum_{\ell=-\infty}^\infty\int \frac{dp_z \, p_\perp dp_\perp}{(2\pi)^2}J_\ell^2(p_\perp r) \nonumber \\
    &\times \left[ \frac{E_p}{2} +T \ln (1-e^{-E_p/T})\right. \nonumber \\ 
    &\left.-\frac{\Omega^2\ell^2}{2T}A_2(E_p) -\frac{\Omega^4\ell^4}{24T^3}A_4(E_p)+\mathcal{O}\left(\Omega^6\right) \right],
    \label{V1expanded}
\end{align}
where
\begin{align}
    A_2(E_p)&\equiv n_B(E_p)\left[1+n_B(E_p)\right], \nonumber \\
    A_4(E_p)&\equiv n_B(E_p)\left[ 1+n_B(E_p) \right]\nonumber \\
    &\times \left[ 1+6n_B(E_p)+6n_B(E_p)^2\right)].
\end{align}

We perform the sum over the angular-momentum modes, using the identities from Eq.~(\ref{sumangularmodes}) and obtain
\begin{align}
   V_1(v;T,\Omega)&=\frac{2}{R^2}\int_0^R rdr \int \frac{dp_z \, p_\perp dp_\perp}{(2\pi)^2} \left[ \frac{E_p}{2} \right. \nonumber \\  
   &\left. +T \ln (1-e^{-E_p/T})-\frac{\Omega^2(p_\perp r)^2}{4T}A_2(E_p) \right. \nonumber \\
   &\left. -\frac{\Omega^4}{T^3}\left( \frac{(p_\perp r)^2}{48}+\frac{(p_\perp r)^4}{64} \right) A_4(E_p) \right]
\end{align}
The radial integration is analogous to the self-energy computation. Then, using Eq.~(\ref{intradial}), the 1-loop potential finally becomes
\begin{align}
    V_1(v;T,\Omega)&= \int \frac{dp_z \, p_\perp dp_\perp}{(2\pi)^2} \left[ \frac{E_p}{2}+T\ln(1-e^{-E_p/T}) \right. \nonumber \\
    &\left. -\frac{\Omega^2(p_\perp R)^2}{8T}A_2(E_p)\right. \nonumber \\
    &\left.-\frac{\Omega^4}{T^3}\left( \frac{(p_\perp R)^2}{96}+\frac{(p_\perp R)^4}{192} \right)A_4(E_p)\right].
    \label{V1final}
\end{align}

Having obtained the one-loop effective potential and the corresponding scalar self-energy, we now incorporate screening effects through the resummation of the ring diagrams. As discussed above, this resummation accounts for the modification of the scalar propagation induced by the medium and becomes particularly relevant in the vicinity of the phase transition, where infrared effects are enhanced.

The ring resummation amounts to dressing the scalar propagator through successive insertions of the self-energy. As a result, the field-dependent squared mass appearing in the infrared-sensitive bosonic sector is replaced according to
\begin{equation}
    m^2 \longrightarrow m^2+\Pi(T,\Omega).
    \label{ScreenedMass}
\end{equation}
In this way, the collective response of the medium is incorporated into the scalar propagation through the self-energy calculated above.

This resummation is particularly important in the region where the field-dependent squared mass $m^2$ becomes negative. At one-loop order, such values can lead to infrared-sensitive and nonanalytic contributions to the effective potential. The replacement $m^2\rightarrow m^2+\Pi(T,\Omega)$ reorganizes these contributions by including the screening effects generated by the medium, thereby providing an improved description of the effective potential in the vicinity of symmetry restoration.

The effective potential including the ring contribution is therefore written as
\begin{equation}
    V_{\rm eff}(v;T,\Omega)=V_{\rm tree}(v)+V_1(v;T,\Omega)+V_{\rm ring}(v;T,\Omega),
    \label{VeffFull}
\end{equation}
where $V_{\rm ring}$ contains the contributions generated by the
resummation of the scalar self-energy insertions. For the implementation of the ring resummation, we retain only the medium-dependent contribution to the scalar self-energy. The vacuum term identified in Eq.~(\ref{selfenergyfinalexpression}) is therefore subtracted from $\Pi(T,\Omega)$, so that the self-energy entering the screened mass accounts exclusively for the screening effects induced by the thermal and rotating medium. Hence, from this point onward, $\Pi(T,\Omega)$ denotes the medium contribution,
\begin{align}
\Pi(T,\Omega)=&\,3\lambda\int\frac{dp_z\,p_\perp dp_\perp}{(2\pi)^2}
\frac{1}{2E_p} \Bigg[ 2n_B(E_p) \nonumber \\
&+\frac{\Omega^2(p_\perp R)^2}{4T^2}B_2(E_p) \nonumber\\
&+\frac{\Omega^4}{T^4}\left(\frac{(p_\perp R)^2}{48}+\frac{(p_\perp R)^4}{96}\right)B_4(E_p)\Bigg].
\label{PiMedium}
\end{align}
Notice that, while only the medium contribution to the self-energy is used in the screened mass, the vacuum contribution to the one-loop effective potential must still be retained. The latter is ultraviolet divergent and therefore requires regularization and renormalization. With this prescription, the explicit expression for the effective
potential becomes
\begin{align}
    V_{\rm eff}(v;T,\Omega)&= -\frac{a^2}{2}v^2+\frac{\lambda}{4}v^4+\int \frac{dp_z \, p_\perp dp_\perp}{(2\pi)^2}\nonumber \\
    &\times \left[ \frac{\tilde{E}_p}{2}+T\ln\left(1-e^{-\tilde{E}_p/T}\right) \right. \nonumber \\
    &\left. -\frac{\Omega^2(p_\perp R)^2}{8T}A_2\left(\tilde{E}_p\right) \right. \nonumber \\
    &\left. -\frac{\Omega^4}{T^3}\left( \frac{(p_\perp R)^2}{96}+\frac{(p_\perp R)^4}{192} \right)A_4\left(\tilde{E}_p\right)\right],
    \label{Veffcomplete}
\end{align}
where $\tilde{E}_p\equiv\sqrt{p_z^2+p_\perp^2+m^2+\Pi(T,\Omega)}$. The vacuum contribution is evaluated using dimensional regularization and renormalized within the $\overline{\mathrm{MS}}$ scheme. We thus obtain
\begin{align}
    V_{\rm eff}(v;T,\Omega)&= -\frac{a^2}{2}v^2+\frac{\lambda}{4}v^4\nonumber \\
    &-\frac{(m^2+\Pi(T,\Omega))^2}{64\pi^2}\left[\ln\left(\frac{\mu^2}{m^2+\Pi(T,\Omega)}\right)+\frac{3}{2}\right] \nonumber \\
    &+\int \frac{dp_z \, p_\perp dp_\perp}{(2\pi)^2}\left[ T\ln\left(1-e^{-\tilde{E}_p/T}\right) \right. \nonumber \\
    &\left. -\frac{\Omega^2(p_\perp R)^2}{8T}A_2\left(\tilde{E}_p\right) \right. \nonumber \\
    &\left. -\frac{\Omega^4}{T^3}\left( \frac{(p_\perp R)^2}{96}+\frac{(p_\perp R)^4}{192} \right)A_4\left(\tilde{E}_p\right)\right],
    \label{VeffcompleteRegularized}    
\end{align}
where $\mu$ is the renormalization scale.

\begin{figure}[b]
    \centering
    \includegraphics[scale=0.58]{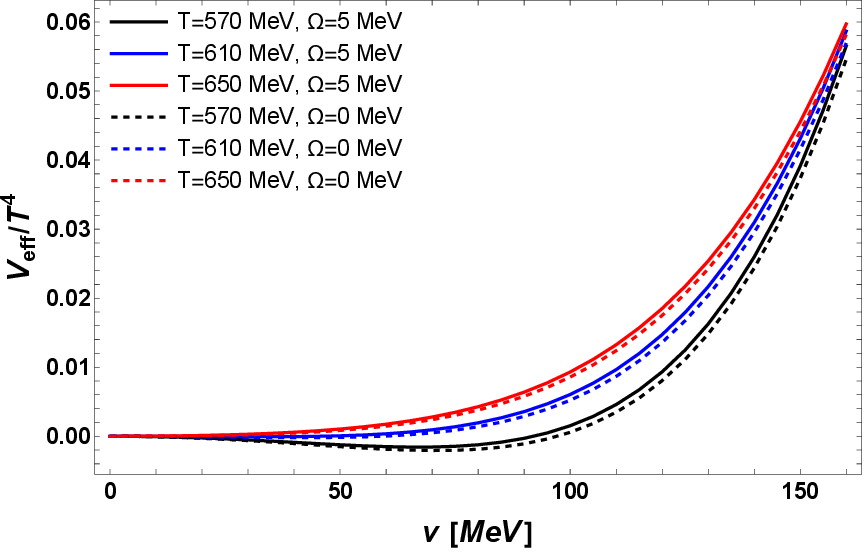}
    \caption{Effective potential normalized by $T^4$ as a function of the order parameter $v$ for $T=570$, $610$, and $650~{\rm MeV}$. Solid curves correspond to a rotating system with $\Omega=5~{\rm MeV}$ and $R=1/20~{\rm MeV}^{-1}$, while dashed curves show the corresponding nonrotating results, $\Omega=0$.}
    \label{fig1}
\end{figure}

Equation~(\ref{VeffcompleteRegularized}) constitutes the effective potential that will be used throughout the remainder of this work. It incorporates the vacuum and medium contributions obtained from the rotating scalar propagator, together with the screening effects encoded in the medium-dependent self-energy $\Pi(T,\Omega)$. Notice that temperature and rotation enter the effective potential not only through the thermal and vortical terms explicitly displayed in Eq.~\eqref{VeffcompleteRegularized}, but also through the replacement $m^2\to m^2+\Pi(T,\Omega)$. The latter accounts for the collective response of the medium and becomes particularly relevant in the vicinity of the phase transition.

Having established the effective potential including the ring resummation, we are now in a position to investigate the evolution of its minima as the temperature and angular velocity are varied. In the next section, we use Eq.~\eqref{VeffcompleteRegularized} to analyze the restoration of the spontaneously broken symmetry and determine how global rotation and screening modify the phase-transition pattern of the scalar theory.

\section{Symmetry Restoration under Global Rotation \label{Sec4}}

We now use the effective potential derived in the previous section to study the restoration of the spontaneously broken $\mathbb{Z}_2$ symmetry at finite temperature and global rotation. Besides exploring the effects of rotation on the symmetry-restoration pattern, this analysis provides an additional consistency check of the rotating scalar propagator developed in this work. In particular, we examine whether the resulting effective potential exhibits the expected physical behavior as the temperature is increased, and subsequently investigate how this behavior is modified by global rotation.

The model contains two free parameters, $\lambda$ and $a$, which we choose as $\lambda=28$ and $a=212$ MeV. These values completely determine the vacuum properties of the model. For the renormalization scale, we take $\mu=1~{\rm GeV}$, consistently with the ultraviolet cutoff employed in the numerical evaluation. Unless otherwise stated, these parameter values are used throughout the following analysis.

We first examine the temperature dependence of the effective potential. Figure~\ref{fig1} shows the effective potential normalized by $T^4$ as a function of the order parameter $v$ for three representative temperatures. The solid curves correspond to a rotating medium with $\Omega=5~{\rm MeV}$ and $R=1/20~{\rm MeV}^{-1}$, while the dashed curves show the corresponding results in the nonrotating case, $\Omega=0$. The latter provides a useful consistency check, since the expected thermal behavior is recovered: as the temperature increases, the minimum of the effective potential moves toward the origin, signaling the restoration of the spontaneously broken $\mathbb{Z}_2$ symmetry. When rotation is included, the same restoration pattern is observed, but it occurs more readily, indicating that global rotation catalyzes symmetry restoration.

\begin{figure}[b]
    \centering
    \includegraphics[scale=0.58]{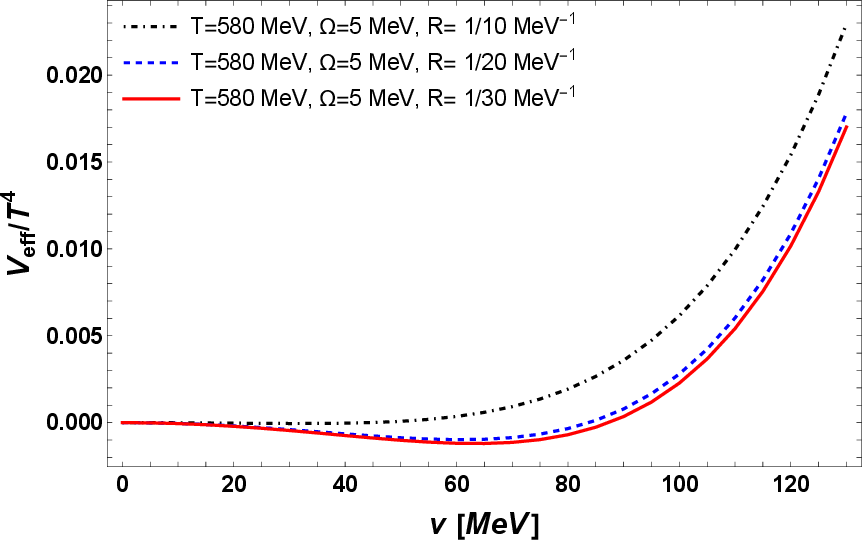}
    \caption{Effective potential normalized by $T^4$ as a function of the order parameter $v$ for different transverse sizes of the rotating system at fixed $T=580~{\rm MeV}$ and $\Omega=5~{\rm MeV}$. Dash-dotted black, dashed blue, and solid red curves correspond to $R=1/10$, $1/20$, and $1/30~{\rm MeV}^{-1}$, respectively.}
    \label{fig2}
\end{figure}

We next examine the dependence of the effective potential on the size of the rotating system. Figure~\ref{fig2} shows $V_{\rm eff}/T^4$ as a function of the order parameter $v$ for different values of the cylinder radius $R$, while keeping the temperature and angular velocity fixed. As $R$ increases, the minimum of the effective potential moves toward the origin, indicating that symmetry restoration is favored in larger rotating systems. Since the tangential velocity in rigid rotation increases with the distance from the rotation axis, $v_{\rm tan}(r)=\Omega r$, increasing $R$ at fixed $\Omega$ also increases the maximum tangential velocity, $v_{\rm tan}^{\rm max}=\Omega R$. Therefore, the observed behavior can be understood as an enhancement of the rotational effects as the transverse size of the system increases.

Taken together, these results show that, within the $\lambda\phi^4$ model considered here, the effective potential constructed from the rotating scalar propagator exhibits the expected thermal symmetry-restoration behavior. In this model, global rotation favors the restoration of the spontaneously broken $\mathbb{Z}_2$ symmetry, and this effect becomes more pronounced as the transverse size of the rotating system increases. At fixed angular velocity, a larger radius also corresponds to a larger maximum tangential velocity, $v_{\rm tan}^{\rm max}=\Omega R$, thereby enhancing the effects associated with rotation. Besides illustrating these physical consequences of global rotation within this model, the recovery of the expected thermal behavior provides a further consistency check of the propagator developed in this work.

\section{Summary and Conclusions \label{Sec5}}

In this work, we have revisited the propagation of a scalar field in a thermal medium under global rotation. The main goal has been to derive a consistent form of the scalar propagator that properly accounts for the spatial dependence introduced by rotation and to examine its use in perturbative calculations. As a controlled framework for this purpose, we considered the $\lambda\phi^4$ model at finite temperature and global rotation, where the propagator was used to compute the one-loop self-energy and, subsequently, the effective potential including screening effects through the resummation of ring diagrams.

The main result of this work is the structure of the rotating scalar propagator itself. By formulating the problem without restricting the description to the comoving frame, the propagator retains an explicit dependence on the radial coordinates $r$ and $r'$, reflecting the loss of translational invariance in the transverse plane induced by global rotation. Consequently, the propagator cannot, in general, be expressed solely in terms of the coordinate difference $x-x'$. This feature has important consequences for perturbative calculations in a rotating medium. In particular, for diagrams involving propagators connecting two or more different spacetime points, the calculation cannot, in general, be formulated exclusively in momentum space from the outset. Instead, the coordinate dependence of the propagators must be retained and the corresponding position-space integrations have to be carried out explicitly. Therefore, higher-order diagrams with several vertices are expected to require a more elaborate treatment than in translationally invariant systems.

Several consistency checks support the form of the propagator derived here. First, in the nonrotating limit, $\Omega\to0$, the explicit radial dependence combines through the Bessel-function addition theorem and the standard free Feynman propagator is exactly recovered, without the need to impose additional relations among the spacetime coordinates. Second, when the rotating propagator is used in the one-loop tadpole diagram, the Matsubara sum correctly generates the vacuum contribution together with the Bose-Einstein medium contributions. Moreover, taking $\Omega\to0$ in the resulting self-energy reproduces the standard finite-temperature expression for a neutral scalar field. These results provide nontrivial checks not only of the nonrotating limit of the propagator, but also of its implementation in an actual perturbative calculation.

To further explore its consequences, we used the resulting self-energy to construct the effective potential of the $\lambda\phi^4$ model. For the analytical and numerical treatment of the rotational corrections, we considered the regime $\Omega/T\ll1$ and retained terms up to $\mathcal{O}(\Omega^4)$. Screening effects were incorporated through the resummation of ring diagrams, with the medium contribution to the self-energy entering the screened scalar mass. The vacuum contribution to the one-loop effective potential was retained and renormalized using dimensional regularization within the $\overline{\rm MS}$ scheme.

The numerical analysis provides a further physical consistency check. Within the $\lambda\phi^4$ model considered here, the effective potential exhibits the expected thermal restoration of the spontaneously broken $\mathbb{Z}_2$ symmetry as the temperature increases. In addition, our results indicate that global rotation favors symmetry restoration in this model. At fixed temperature and angular velocity, this effect becomes more pronounced as the transverse size of the rotating system increases. Since the maximum tangential velocity is $v_{\rm tan}^{\rm max}=\Omega R$, increasing the radius at fixed $\Omega$ enhances the rotational effects, in
agreement with the explicit dependence on $R$ found in the effective potential.

Altogether, the recovery of the free propagator in the nonrotating limit, the correct vacuum and medium structure of the one-loop self-energy, and the expected thermal behavior of the effective potential provide complementary consistency checks of the rotating scalar propagator developed in this work. At the same time, its explicit spatial dependence points to an important consequence for future applications. While the tadpole calculation considered here involves the propagator at coincident points, more general perturbative diagrams will contain propagators connecting different spacetime points, for which the radial dependence cannot be avoided. We expect that an analogous structure will emerge for fermionic and gauge fields under global rotation, leading to propagators involving both coordinate- and momentum-space dependence. If this expectation is confirmed, perturbative calculations involving these fields will likewise require a careful treatment of their position-space structure. The extension of the present formulation to fermionic and gauge fields, together with its implementation in more general perturbative calculations, is currently under investigation.

\begin{acknowledgments}
Support for this work was received in part by the Secretaria de Ciencia, Humanidades, Tecnología e Innovación Grant No. CBF-2025-G-1718. RZ acknowledges support from ANID/CONICYT FONDECYT Regular (Chile) under Grant No. 1241436.
\end{acknowledgments}

\bibliography{mybibliography}% Produces the bibliography via BibTeX.

\end{document}